\documentclass[
 aip,
 jmp,
 amsmath,amssymb,
 reprint,
]{revtex4-1}

\usepackage{graphicx}
\usepackage{dcolumn}
\usepackage{bm}

\begin{document}
\preprint{}
\title{Iterative resource allocation based on propagation feature of node for identifying the influential nodes}
\author{Lin-Feng Zhong}
\affiliation{Web Science Center, University of Electronic Science and Technology, Chengdu 611731, P. R. China}
\affiliation{Big Data Research Center, University of Electronic Science and Technology, Chengdu 611731, P. R. China}
\author{Jian-Guo Liu}
\email[]{liujg004@ustc.edu.cn}
\affiliation{Research Center of Complex Systems Science, University of Shanghai for Science and Technology, Shanghai 200093, P. R. China}
\author{Ming-Sheng Shang}
\email[]{shang.mingsheng@gmail.com}
\affiliation{Web Science Center, University of Electronic Science and Technology, Chengdu 611731, P. R. China}
\affiliation{Big Data Research Center, University of Electronic Science and Technology, Chengdu 611731, P. R. China}

\date{\today}

\begin{abstract}
The Identification of the influential nodes in networks is one of the most promising domains. In this paper, we present an improved iterative resource allocation (IIRA) method by considering the centrality information of neighbors and the influence of spreading rate for a target node. Comparing with the results of the Susceptible Infected Recovered (SIR) model for four real networks, the IIRA method could identify influential nodes more accurately than the tradition IRA method. Specially, in the Erd\"{o}s network, the Kendall's tau could be enhanced 23\% when the spreading rate is 0.12. In the Protein network, the Kendall's tau could be enhanced 24\% when the spreading rate is 0.08.
\end{abstract}
\maketitle
\section{Introduction}
Spreading is a ubiquitous phenomena in nature. A lot of activities can be seen as spreading in society\cite{GINSBERG2008,WANG2009,CENTOLA2010,Zhou2006}. In the past few years, the spreading in complex networks is concerned more and more with its great theoretical significance and remarkably practical value, that is, epidemic controlling\cite{Albert2002,PASTOR2001,KEEELING2008}, information dissemination\cite{LIU2007} and viral marketing. One of the fundamental problems is to identify the influential nodes in the networks. The knowledge of the node's spreading ability shows new insights for applications such as identifying influential nodes\cite{ARAL2012,LIU2013-A,IYER2013,BELLINGERI2014}, designing efficient methods to either hinder epidemic spreading or accelerate information dissemination.

Recently, there are a lot of centrality methods\cite{LIU2013} have been applied to identify the influential nodes in complex networks, including degree, eigenvector centrality\cite{BORGATTI2005}, closeness centrality\cite{SABIDUSSI1966} and $k$-shell decomposition\cite{KITSAK2010}. The degree centrality is based on the number of neighbors connected with a node. Chen {\it et al}.\cite{CHEN2012} defined a local centrality based on the degree information of nearest neighbors and the second nearest neighbors. Poulin {\it et al}.\cite{POULIN2000} proposed cumulated nomination centrality based on the iterative method for solving the feature vector mapping. Zhang {\it et al}.\cite{ZHANG2011} proposed a multiscale measurement by considering the interactions from all the paths a node is involved. Kitsak {\it et al}.\cite{KITSAK2010} found that the most influential nodes are those located within the core of the network by decomposing a network with the $k$-shell decomposition method. By taking into account the neighbors' $k$-core values, Lin {\it et al}.\cite{LIN2014} proposed an improved neighbors' $k$-core (INK) method to identify the influential nodes with the largest $k$-core values. And by considering the $k$-shell decomposition method and resource iteration, Ma {\it et al}.\cite{MA2014} proposed an improved method to identify the influential nodes. In directed networks, some methods are also designed to identify the influential spreaders namely LeaderRank\cite{LV2011}, which outperforms the well-known PageRank method in both effectiveness and robustness.

\begin{figure}[ht]
\center\scalebox{0.38}[0.38]{\includegraphics{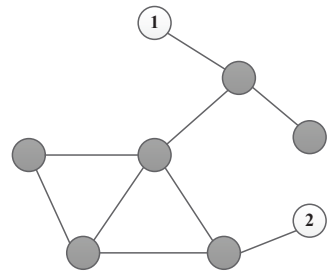}}
\caption{(Color online) An example network consists of 8 nodes and 9 edges. We employ the SIR model to simulate the spreading process for these nodes. Initially, only one node is selected to be infected. For each node, the node spreading influence is defined as the number of the infected nodes, then the spreading influence of nodes 1, 2 is 1.079, 1.083 respectively. The result is obtained by averaging over 10000 independent runs and 3 time steps when the spreading rate $\beta$ is 0.07. In the IRA method, the value of nodes 1, 2 which generated by the IRA method is 0.33, 0.29 respectively.}\label{Fig1}
\end{figure}
These existing methods mainly consider the significance of node, but the node influence is also affected by the importance of its neighbors. Based on this idea, Ren{\it et al}.\cite{REN2014} proposed an iterative resource allocation (IRA) method to identify influential nodes. However, the number of infected neighbor also affect the resource allocation. Take Fig. 1 as example, the IRA method can not distinguish the influence of the nodes 1 and 2 accurately. Therefore, we argue that either the number of neighbors and the spreading rate may affect the property of the target nodes simultaneously. Inspired by this idea, we present an improved iterative resource allocation method (IIRA), where the resource allocation of the target node may adjust by the spreading rate. Then, comparing with the Susceptible Infected Recovered (SIR) spreading process\cite{BARRAT2008,NEWMAN2002} for four real networks, the results show that the ranking list generated by our method could identify influential nodes more accurately than the one generated by the IRA method.

\section{the IIRA method}
The traditional IRA method supposes that the node influence is determined by neighbors' centralities, such as the degree centrality, $k$-shell method, closeness centrality, betweenness centrality and so on. There is a tunable parameter to nonlinearly adjust the weight of the centrality. The IIRA method also requires the traditional node centralities as the input, such as degree centrality, $k$-shell method\cite{KITSAK2010}, closeness centrality\cite{SABIDUSSI1966} and eigenvector centrality\cite{BORGATTI2005}. In the IIRA method, we argue that each node has a initial resource in the initial state, and the spreading rate could adjust the allocation of the resource which determined by the neighbors' centralities. After a certain number of iterations, the resource of each node will approach a steady state and the final amount of resource in each node will be used to identify the influential nodes. The IIRA method can be described as follows in detail.

We consider that an undirected network $G=(N,E)$ with $N$ nodes and $E$ edges could be described by an adjacent matrix $\Omega=\{\delta_{ij}\}\in R^{n,n}$ where $\delta_{ij}=1$ if node $i$ is connected by node $j$, and $\delta_{ij}=0$ otherwise. At each iteration step, each node could allocate the resource determined by the node centrality to it's neighbors, and the resource allocation is adjusted by the spreading rate. Let $\Gamma(i)$ be set of node $i$'s neighbors, the resource of node $i$ obtained by resource allocation can be expressed as follows:

\begin{equation}\label{equation1}
\begin{aligned}
 {\rm I}_{i}(t+1)& = \sum_{j\in \Gamma(i)}R_{j\rightarrow i}(t+1) \\
 & =\sum_{j\in \Gamma(i)}\left[\psi_{i}\theta_{i}\left(\sum_{u\in \Gamma(j)}\theta_{u}\right)^{-1}{\delta_{ij}}\right]{\rm I}_{j}(t),\\
\end{aligned}
\end{equation}
where $R_{j\rightarrow i}(t+1)$ represents that the node $j$ allocates its resource to node $i$ at step $t+1$, $I_{j}(t)$ represents the node $j$'s resource at step $t$, $\theta_{i}$ is the centrality of the node $i$, $\psi_{i}$ reflects the influence of change spreading rate of the node $i$, $\beta$ is the spreading rate, and the $k_{i}$ is the degree of node $i$, $k_{i}$ can be expressed as

\begin{equation}\label{equation2}
k_{i}=\sum_{j\in G}\delta_{ij},
\end{equation}
Equation (1) can be written in matrix form:

\begin{equation}\label{equation3}
{\rm I}(t+1)={\bf A}{\rm I}(t)=
\left(
\begin{array}{ccc}
a_{11}&\ldots&a_{1n}\\
\vdots&\ddots&\vdots\\
a_{n1}&\ldots&a_{nn}\\
\end{array}
\right)
\left(
\begin{array}{c}
I_{1}(t)\\
\vdots\\
I_{n}(t)\\
\end{array}
\right),
\end{equation}
where the element $a_{ij}$ of matrix ${\bf A}$ is given by

\begin{equation}\label{equation4}
a_{ij}=[1-(1-\beta)^{k_{i}}]\theta_{i}\left(\sum_{u\in \Gamma(j)}\theta_{u}\right)^{-1}\delta_{ij}.
\end{equation}

We assume that each node has a initial resource(${\rm I}(0)=(1,1,\cdots,1)^{T}$). The ultimate resource of each node will reach a new state(${\rm I}(t)={\bf A}{\rm I}(t-1)={\bf A}^{t}{\rm I}(0)$). According to the Gershg\"{o}rin disk theorem\cite{HORN1985}, the spectral radius $\rho(A)$ of matrix {\bf A} is no larger than 1, and the vector ${\rm I}(t)$ will converge to a steady value after t times iterations. However, the t is not infinite. Then the ranking list generated by the ${\rm I}(t)$ can identify the influential nodes.

\begin{figure}[ht]
\center\scalebox{0.38}[0.38]{\includegraphics{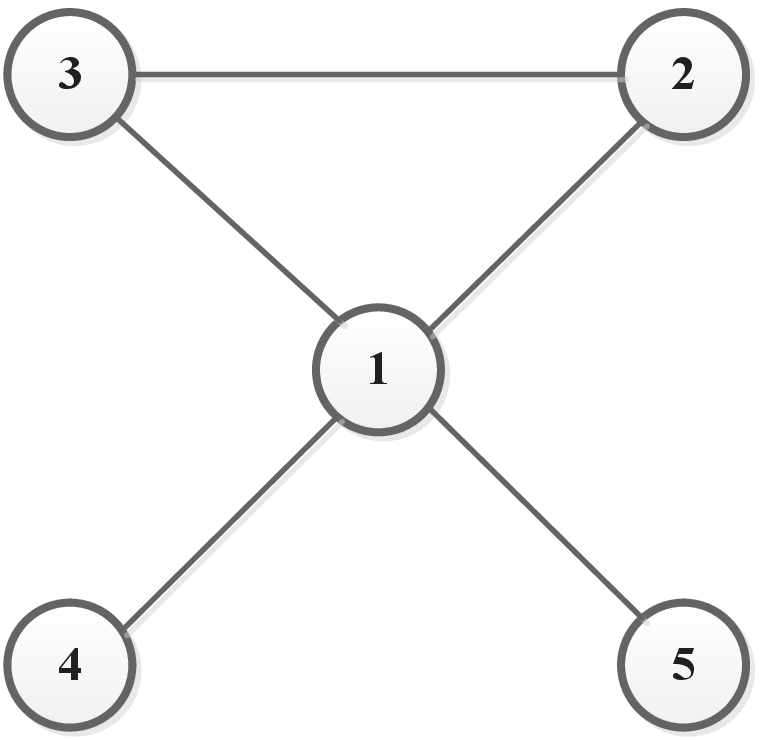}}
\caption{(Color online) Example of networks with node $N=5$. the node label is 1, 2, 3, 4, 5, respectively. $ks_{1,2,3}=2$, $ks_{3,4}=1$.}\label{Fig2}
\end{figure}

For example, a network with 5 nodes and 5 edges is shown in Fig. 2. The initial resource of each node equals 1(${\rm I}(0)=(1,1,1,1,1)^{T}$). We take $k$-shell as the node centrality($\theta=ks$ as an example) and set spreading rate $\beta=0.2$. The resource allocation matrix is as follows:

\begin{equation}\label{equation5}
{\bf A}=
\left(
\begin{array}{ccccc}
0&0.29&0.29&0.59&0.59\\
0.12&0&0.18&0&0\\
0.12&0.18&0&0&0\\
0.03&0&0&0&0\\
0.03&0&0&0&0\\
\end{array}
\right),
\end{equation}
To make the final resource allocation value convergent,we set iterations times $t=50$, ${\rm I}(50)={\bf A}^{50}{\rm I}(0)=[8.19e-20, 4.32e-20, 4.32e-20, 6.7e-21, 6.7e-21]^{T}$. The result shows that the IIRA method can estimate the influence of the target node.

\section{Experiment Results}
\subsection{Data description}
In this paper, we evaluate the performance of the IIRA method in four networks including the Erd\"{o}s, US air line, Email and Protein networks. The Erd\"{o}s network\cite{BARABAS2002} is a scientific collaboration networks. Each node represents the scientist whose Erd\"{o}s number is 1 and the edge represents the cooperative connection between each pair of scientists. The US air line network\cite{BATAGELJ1998} is the part of air traffic lines in America. Each node represents the city and the edge represents the airline between the city and the other city. The Email network\cite{LIU2013-A} is the network that each node represents different people including researchers, technicians, managers, administrators and graduate students. The edge between the two nodes indicates that these people keep a communication relation with each other. The Protein network\cite{SUN2003} represents the interaction between the various proteins.

The statistical properties of four real networks are shown in Table I, including the number of nodes $N$, the number of edges $E$, the average degree $\langle k \rangle$ and the epidemic threshold $\beta_{c}$.

\begin{table}[ht]\caption{Basic statistical features of the Erd\"{o}s, US air line, Email and Protein networks, including the number of nodes $N$, edges $E$, the average degree $\langle k \rangle$, the spreading threshold $\beta_{c}$.}
\begin{center}
\begin{tabular} {l r r r r}
  \hline \hline
   Network        &  $N$    &   $E$     &  $\langle k \rangle$   & $\beta_{c}$ \\ \hline
   Erd\"{o}s      &  474    &   1639   &         6.916          &     0.055     \\
   US air line    &  332    &   2126   &         12.807         &     0.021     \\
   Email          &  1133   &   5451   &         9.622          &     0.050     \\
   Protein        &  2284   &   6646   &         5.820          &     0.052     \\
   \hline \hline
    \end{tabular}
\end{center}
\end{table}

\subsection{Measurement}
In this paper, we use the SIR model\cite{BARRAT2008,NEWMAN2002} to examine the node spreading influence. In such a system, there are three compartments: (i) Susceptible individuals represent the number of individuals susceptible to (not yet infected) the disease; (ii) Infected individuals represent individuals who have been infected and are able to spread the disease to susceptible individuals; (iii) Recovered individuals represent individuals who have been recovered and will never be infected again. At each time step, for each infected node, one randomly selected susceptible neighbor gets infected with the spreading rate $\beta$, and the infected node would recover in one time step. The number of infections $X$, generated by the initially-infected node including the initially-infected is denoted as its spreading influence, where $\beta$ is the spreading rate in the SIR model. The number of infections $X$ is obtained by averaging over 20000 independent runs and the time step T is set as 5.

To check the performance of the IIRA method, the Kendall's tau\cite{KENDALL1938} $\tau$ is introduced to measure the correlation of the node spreading influence with the IIRA method and the IRA method. Kendall's tau $\tau$\ is used to measure the correlation between two ranking list. The Kendall's tau $\tau$\ value lies in [-1,1], and the increasing values imply the method can identify the influential nodes more accurately. The Kendall's tau $\tau$\ is defined as

\begin{equation}\label{equation6}
\tau=\frac{2}{N(N-1)}\sum_{i<j}{\rm{sgn}}[(x_{i}-x_{j})(y_{i}-y_{j})],
\end{equation}
where $N$ is the number of nodes in a network, $x_{i}$ is the spreading influence of node $i$, $y_{i}$ are the values of the IIRA method which generated by the degree, closeness, $k$-shell and eigenvector centralities. The $\rm{sgn}$($x$) is a piecewise function, when $x>0$, $\rm{sgn}$($x$)$=+1$; $x<0$, $\rm{sgn}$($x$)$=-1$; When $x=0$, $\rm{sgn}$($x$)$=0$.

\subsection{Simulation results}
In this section, we first verify the effectiveness of the IIRA method. Figure 3 shows Kendall's tau $\tau$ of the IIRA method when the iteration times t changes from 1 to 100. We can find that each Kendall's tau $\tau$
\begin{figure}[ht]
\center\scalebox{0.35}[0.35]{\includegraphics{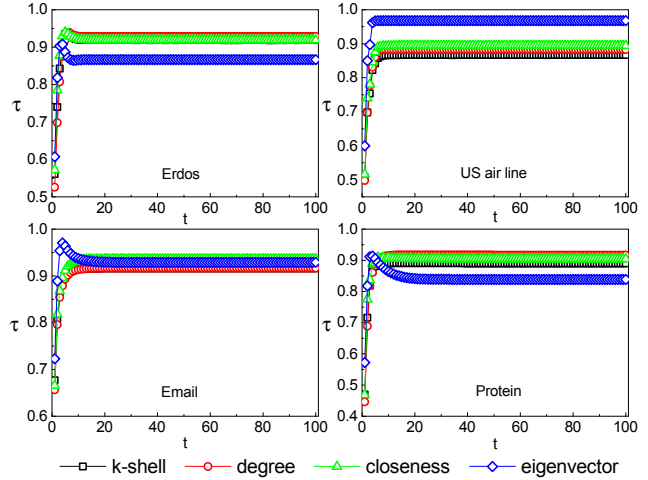}}
\caption{(Color online) The Kendall's tau $\tau$ obtained by comparing the ranking list generated by the SIR spreading process and the ranking lists generated by the $k$-shell (squares), degree (circles), closeness (triangles) and eigenvector (diamonds) centralities. The iteration times $t$ lies $(0,100]$, and the spreading rate $\beta=0.05$. }\label{Fig3}
\end{figure}
gradually achieve stability with increasing of t times iterations. However, in the four networks, each t is different when the Kendall¡¯s tau $\tau$ achieve stability. But we can see that every Kendall's tau $\tau$ approach a steady value when the iterations times t is no larger than 50. We think that the IIRA method can identify the influential nodes effectively when the iterations times $t=50$. So we set the iteration times of the IIRA method $t=50$.

And then, we evaluate the performance of the IIRA method in the four networks. We use relatively small values of $\beta$ in SIR model, namely $\beta\in (0,0.2]$. Figure 4 shows Kendall's tau $\tau$ of the IIRA method where the ranking lists are generated by the $k$-shell, degree, closeness and eigenvector centralities. From which, one can find that there are different performances for the different centralities. For instance, in the Erd\"{o}s network, when the spreading rate $\beta$ is lower than 0.08, the Kendall's tau $\tau$ generated by the eigenvector centrality is lower than the ones generated by other centralities. However, when the spreading rate $\beta$ is larger than 0.08, the Kendall's tau $\tau$ generated by the eigenvector centrality is much larger than other centralities. It means that the ranking list generated by the eigenvector centrality could identify the
\begin{figure}[ht]
\center\scalebox{0.34}[0.34]{\includegraphics{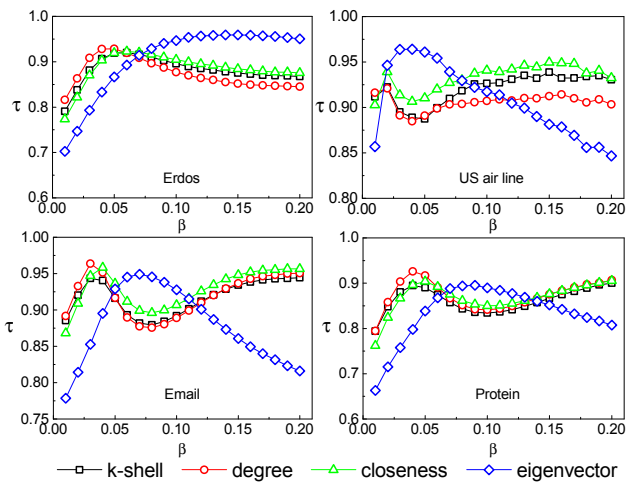}}
\caption{(Color online) The Kendall's tau $\tau$ obtained by comparing the ranking list generated by the SIR spreading process and the ranking lists generated by the $k$-shell (squares), degree (circles), closeness (triangles) and eigenvector (diamonds) centralities. The results are averaged over 20000 independent runs with different spreading rate $\beta$.}\label{Fig4}
\end{figure}
influential nodes more accurately than other centralities. In the US air line network, the Kendall's tau $\tau$ of the IIRA method where the ranking list generated by the eigenvector centrality is much larger than ones generated by other centralities when the spreading rate $\beta$ lies between 0.02 and 0.08. And in the Email network, the Kendall's tau $\tau$ of the IIRA method where the ranking list generated by the eigenvector centrality is much larger than ones generated by other centralities when the spreading rate $\beta$ lies between 0.06 and 0.1. The same phenomena could be found for the Protein network when the spreading rate $\beta$ lies between 0.07 and 0.13.

Figure 4 also shows that, in the four networks, the Kendall's tau $\tau$ of the IIRA method where the ranking list generated by the eigenvector centrality firstly increase and then decrease with the increasing of the spreading rate $\beta$. In the Erd\"{o}s, Email and Protein networks, the Kendall's tau $\tau$ of the IIRA method where the ranking lists generated by the $k$-shell, degree and closeness centralities firstly increase and then decrease, and will increase again after the drop. Particularly, the Kendall's tau $\tau$ generated by the $k$-shell, degree and closeness centralities will increase again after the drop in the US air line, Email and Protein networks.

\begin{figure}[ht]
\center\scalebox{0.34}[0.34]{\includegraphics{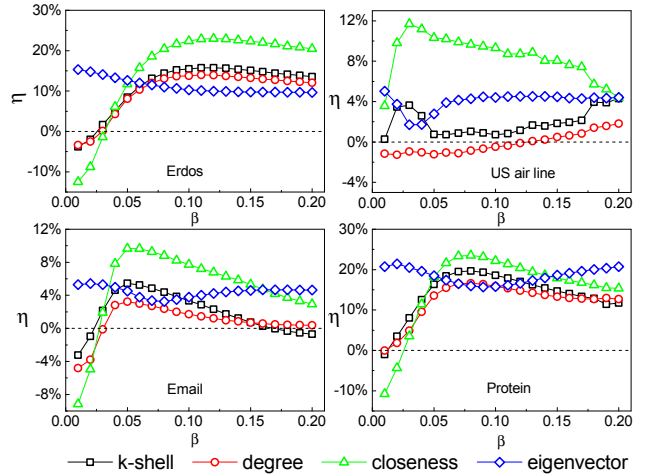}}
\caption{(Color online) The improved ratio $\eta$ of different spreading rates $\beta$ with different centralities including $k$-shell, degree, closeness and eigenvector centralities in the Erd\"{o}s, US air line, Email and Protein networks.}\label{Fig5}
\end{figure}

Figure 5 reports the improved ratio $\eta$ of the Kendall's tau $\tau$ generated by the IIRA method comparing with the results of the IRA method. The improved ratio $\eta$ is defined as
\begin{equation}\label{equation7}
\eta=\frac{\tau^{new}-\tau^{0}}{\tau^{0}},
\end{equation}
where $\tau^{new}$ is the Kendall's tau $\tau$ of the IIRA method by considering the degree, $k$-shell, closeness and eigenvector centralities, and $\tau^{0}$ is the Kendall's tau $\tau$ obtained by the IRA method. Clearly, $\eta>0$ indicates an advantage of the IIRA method. The improved ratios in $\tau$ for the IRA method where the ranking lists generated by the $k$-shell, degree, closeness and eigenvector centralities with different spreading rate $\beta$ are shown in Fig. 5. In the Erd\"{o}s network, the largest improved ratio $\eta$ generated by the closeness centrality could reach 22\% when the spreading rate $\beta$ is 0.12. In the US air line network, the largest $\eta$ generated by the closeness centrality could reach 12\% when the spreading rate $\beta$ is 0.03. The same phenomena could be found for the Email and Protein networks, the improved ratio $\eta$ generated by the closeness centrality could reach the largest value when the spreading rate $\beta$ is 0.05 and 0.08 respectively. In the Erd\"{o}s and Protein networks, when the spreading rate $\beta$ lies between 0.04 and 0.2, the improved ratios $\eta$ generated by the four kinds of centralities are larger than 0 which indicates that the IIRA method could identify the influential nodes more accurately than the IRA method. In the US air line network, the improved ratios $\eta$ generated by the closeness and eigenvector centralities are larger than 0. The improved ratio $\eta$ generated by the closeness centrality is larger than the one generated by the eigenvector centrality, it means that the ranking list of the IIRA method generated by the closeness centrality is more accurate than the one generated by the eigenvector centrality.

Furthermore, we find the dependence of improved ratio $\eta$ on the spreading rate $\beta$ in Fig. 5. In the Erd\"{o}s, Email and Protein networks, we find that the improved ratio $\eta$ generated by the $k$-shell centrality has two distinct trends with the increasing of the spreading rate $\beta$. The improved ratio $\eta$ firstly increase and then decrease gradually. There are the same trends for the degree and closeness centralities. In the Erd\"{o}s and Protein networks, the improved ratios $\eta$ generated by the $k$-shell, degree, closeness, eigenvector centralities  are larger than 0 when the spreading rate $\beta$ is larger than 0.05 which indicates that the IIRA method could identify the influential nodes more accurately than the IRA method. In the US air line network, the improved ratio $\eta$ generated by the closeness centrality improve more obviously than the ones generated by other centralities which indicates that the IIRA method generated by the closeness centrality could identify the node spreading influence more accurately than the IRA method. The same phenomena could be found for the Email network, which indicates that the ranking accuracy of the IIRA method generated by the closeness centrality is much better than the IRA method.

\section{conclusion and discussions}
By taking into account the neighbors' resource of the node and the influence of spreading rate for the target node, we present an improved iterative resource allocation (IIRA) method to identify the node spreading influence. The IIRA method considers the infection status of target node, and the probability of infection determined by the degree of target node. And the new resource allocation is determined by these information. It can be applied to many classical centralities such as degree, $k$-shell, closeness, eigenvector centralities. The simulation results show that the performance of the IIRA method can be further improved than the IRA method without adding any other parameters and computation complexity. In the four networks, the performance of the improved ratio $\eta$ generated by the closeness centrality is better than the ones generated by other centralities. Specially, in the Erd\"{o}s network, the largest improved ratio $\eta$ generated by the closeness centrality could reach 22\% when the spreading rate $\beta$ is 0.12. In the Protein network, the largest improved ratio $\eta$ generated by the closeness centrality could reach 24\% when the spreading rate $\beta$ is 0.08. These results show that the IIRA method could identify the influential nodes more accurately than the IRA method.

In the IIRA method, the final resource of a node is not only determined by its neighbors' centralities but also depended on the number of infected neighbors. It would be very interesting to test the modified centrality in other dynamic process. For example, the IIRA method only considers the influence of spreading rate of target node. Whether the influence of target node's neighbors should be considered. It also should be noticed that the IIRA method operation requires iterative process, which is very time-consuming. So, we should find a feasible method to reduce the computational complexity. In addition, interconnected networks have attracted more and more attention recently. How to design a new iterative resource allocation method in these networks may be an interesting and important open problem.

\begin{acknowledgments}
This work is partially supported by the National Natural Science Foundation of China (Grant Nos. 61370150, 91324002, 71371125), the Shanghai Leading Academic Discipline Project of China (No. XTKX2012), MOE Project of Humanities and Social Science (Grant Nos. 14ZR1427800), JGL is supported by the Program for Professor of Special Appointment (Eastern Scholar) at Shanghai Institutions of Higher Learning.
\end{acknowledgments}

\end{document}